\documentclass[12pt]{iopart}

\usepackage{graphicx}
\usepackage{dcolumn}
\usepackage{bm}
\begin{document}
\newcommand{\er}{{\bf e}_{r}}
\newcommand{\ep}{{\bf e}_\varphi}
\newcommand{\et}{{\bf e}_\theta}
\newcommand{\hh}{{\bf H}}
\newcommand{\ee}{{\bf E}}
\newcommand{\ww}{{\bf W}}
\newcommand{\vv}{{\bf v}}
\newcommand{\rr}{{\bf r}}
\newcommand{\eee}{{\bf e}}
\newcommand{\uu}{{\bf u}}
\newcommand{\lom}{{\bf L}}

\newcommand{\rmq}{{\rm q}}
\newcommand{\rmt}{{\rm t}}
\newcommand{\co}{{\rm co}}
\newcommand{\cl}{{\rm cl}}

\newcommand{\epr}{\varepsilon_{r r}}
\newcommand{\mur}{\mu_{r r}}
\newcommand{\alr}{\alpha_{r r}}
\newcommand{\kar}{\kappa_{r r}}
\newcommand{\Om}{\Omega_a^r}
\newcommand{\ve}{\vec{e}}
\newcommand{\vw}{\vec{w}}
\newcommand{\va}{\vec{a}}

\title[]{Theory and optimization of transformation-based quadratic spherical cloaks}

\author{Andrey Novitsky$^{1}$ and Cheng-Wei Qiu$^{2^*,3}$}

\address{$^{1}$Department of Theoretical Physics, Belarusian
State University, Nezavisimosti Avenue 4, 220050 Minsk, Belarus.
Electronic address: andrey.novitsky@tut.by}

\address{$^{2}$Research Laboratory of Electronics, Massachusetts
Institute of Technology, 77 Massachusetts Avenue, Cambridge, MA
02139, USA.}

\address{$^{3}$Department of Electrical and Computer
Engineering, National University of Singapore, 4 Engineering Drive
3, Singapore 117576.}\ead{cwq@mit.edu}

\begin{abstract}
Based on the concept of the cloak generating function, we propose a
numerical method to compute the invisibility performance of various
quadratic spherical cloaks and obtain optimal invisibility results.
A non-ideal discrete model is used to calculate and optimize the
total scattering cross-sections of different profiles of the
generating function. A bell-shaped quadratic spherical cloak is
found to be the best candidate, which is further optimized by
controlling design parameters involved. Such improved invisibility
is steady even when the model is highly discretized.

\end{abstract}

\maketitle

\section{Introduction}
Recently, great progress has been made in both the theory of and
experiments on invisibility cloaks
\cite{Pendry_sci,Leonhardt_sci2006,microwave,groundplane}. Wide
applications have been found in microwave spectrum
\cite{Leonhardt_NJP2006,Miller_DAB,Schurig_OE2006,Nicorovici_OE07,Liang_APL08,Zhao_OE08},
optical regime
\cite{Cai_NP,Cai_OE08,Vanbesien,Xiao_OL,Jenkins,ZhangXiang},
elastodynamics \cite{Milton_NJP,Farhat_PRB}, quantum mechanics
\cite{quantum1,quantum2}, and acoustics
\cite{CTChan_2007,Cummer_2007,Sanchez_NJP,Cummer_2008}. One approach
to achieve an invisibility cloak is to employ the transformation
optics (TO) to allow electromagnetic waves to be directed around the
concealed region and smoothly recovered afterwards. The anisotropic
parameters of such a cloak are derived from the coordinate
transformation. This approach was generalized from the cloaking of
thermal conductivity \cite{Greenleaf_2003} and then widely applied
in many other areas, which provides new approaches to conceal
passive/active objects \cite{ChenHS_2007,ZhangBL_2008} within their
interiors invisible to external illuminations. The fundamental idea
is the invariance of Maxwell's equations under a space-deforming
transformation if the material properties are altered accordingly;
i.e., a specific spatial compression is equivalent to a variation of
the material parameters in the flat space. Based on TO concept, many
efforts have been devoted to the study of 2D cloaks (cylindrical
\cite{cylinder}, elliptical \cite{ellipse}, and arbitrary
cross-section \cite{arbitrary}) due to the simplicity in numerical
simulations. Inspired by the classic spherical cloak
\cite{Pendry_sci}, the expressions of
  electromagnetic fields were explicitly presented in terms of
  spherical Bessel functions by Mie theory
  \cite{ChenHS_2007}. However, this analytical scattering theory for classic spherical cloaks cannot work if the anisotropic ratio (see the original definition in
  \cite{Qiu_PRE2007}) is anything other than that in
  \cite{ChenHS_2007}. Two solutions to overcome this problem were
  proposed: 1) multilayers of alternating isotropic layers \cite{Qiu_PRE2009}; and 2) discrete model of the inhomogeneous
  anisotropic shell, and each layer is radially anisotropic but
  homogeneous \cite{MaHua}. Then, the spherical invisibility cloak is
  near-perfect.
  Another non-TO route to the cloaking in
canonical shape is to use a homogenous anisotropic
\cite{GaoLei_PRE,Yaxian} or isotropic plasmonic \cite{Alu} coating.
However, in this method, the effectiveness and properties of the
cloak depend on the object to be cloaked as well as that its size
has to be sufficiently small compared with the wavelength. Usually,
TO-based spherical cloaks need to know the prescribed transformation
functions first and the required parameters can thus be obtained by
constructing the explicit transformation matrices.

Certainly, those sets of parameters from various generating
functions for quadratic spherical cloaks are ideal, all of which
should give zero scattering theoretically. However, in actual
situations, one has to consider a discrete multilayered model so
that the invisibility performances of different generating functions
distinguish from each other. The general method developed in
\cite{MaHua} is adopted to calculate the far-field scattering. Our
numerical results reveal that the power quadratic bell-shaped cloak
yields the lowest scattering under the same discretization, which is
still pronounced when the ideal cloaking shell is highly
discretized.

This paper is organized as follows. Section 2 proposes the analysis
method which will be used to discretize the cloaking shell into
multiple layers and then compute the far-field diagrams. Section 3
addresses that the bell-shaped profile of the generating function
outperforms the other including the linear one corresponding to the
classic spherical cloak. Section 4 compares different profiles which
give rise to bell-shaped profiles. Section 5 discusses the
optimization of bell-shaped quadratic cloaks where the steady
improvement in invisibility performance is verified.

\section{Scattering Algorithm for Arbitrary Spherical Cloaks in Discretized Model}
In this section, the scattering theory of multilayer anisotropic
spherical particles is provided and applied to study a spherical
cloak with arbitrary transverse and radial parameters. We suppose
that the arbitrary field distribution of the incident monochromatic
wave interacts with the two-layer sphere. The inner sphere is
supposed to be made of isotropic material. The parameters of the
coating depend on the radial coordinate and specify the rotationally
symmetric anisotropy of the form
\begin{eqnarray}
\varepsilon = \varepsilon_r(r) \er \otimes \er + \varepsilon_t(r) I,
\qquad \mu = \mu_r(r) \er \otimes \er + \mu_t(r) I, \label{eps_mu}
\end{eqnarray}
where $\varepsilon_r$ and $\mu_r$ are the radial dielectric
permittivity and magnetic permeability, $\varepsilon_t$ and $\mu_t$
are the transversal material parameters, $I = 1 - \er \otimes \er =
\et \otimes \et + \ep \otimes \ep$ is the projection operator onto
the plane perpendicular to the vector $\er$, unit vectors $\er$,
$\et$, and $\ep$ are the basis vectors of the spherical coordinates.

Using the separation of the variables, the solution of Maxwell's
equations in spherical coordinates ($r$, $\theta$, $\varphi$) can be
presented as
\begin{eqnarray}
\ee (r, \theta, \varphi) = F_{lm}(\theta, \varphi) \ee(r), \nonumber
\\ \hh (r, \theta, \varphi) = F_{lm}(\theta, \varphi) \hh(r),
\label{solution1}
\end{eqnarray}
where the designation $\ee (r)$ means that the components of the
electric field vector depend only on the radial coordinate $r$ as
$E_r(r)$, $E_\theta(r)$, and $E_\varphi(r)$ (however, the vector
itself includes the angle dependence in the basis vectors), the
second rank tensor in three-dimensional space $F_{lm}$ serves to
separate the variables ($l$ and $m$ are the integer numbers). It can
be written as the sum of dyads:
\begin{equation}
F_{lm}=Y_{lm} \er \otimes \er + {\bf X}_{lm} \otimes \et + (\er
\times {\bf X}_{lm}) \otimes \ep. \label{Flm}
\end{equation}
where $Y_{lm}(\theta, \varphi)$ and ${\bf X}_{lm}(\theta, \varphi)$
are the scalar and vector spherical harmonics. Tensor functions
$F_{lm}$ are very useful because of their orthogonality conditions
\begin{equation}
\int_0^\pi \int_0^{2 \pi} F_{l'm'}^+(\theta, \varphi) F_{lm}(\theta,
\varphi) \sin \theta {\rm d} \theta {\rm d} \varphi = {\bf 1}
\delta_{l'l} \delta_{m'm}. \label{orthogF}
\end{equation}
From the commutation of the dielectric permittivity $\varepsilon$
and magnetic permeability $\mu$ tensors with $F_{lm}$ it follows
that the electric and magnetic fields satisfy the system of ordinary
differential equations
\begin{eqnarray}
&&\er^\times \frac{{\rm d} \hh}{{\rm d} r} + \frac{1}{r} \er^\times
\hh - \frac{\rmi \sqrt{l(l+1)}}{r} \ep^\times \hh = - \rmi k_0
\varepsilon \ee, \nonumber \\
&&\er^\times \frac{{\rm d} \ee}{{\rm d} r} + \frac{1}{r} \er^\times
\ee - \frac{\rmi \sqrt{l(l+1)}}{r} \ep^\times \ee = \rmi k_0 \mu
\hh. \label{ODE1}
\end{eqnarray}
where $k_0 = \omega / c$ is the wavenumber in vacuum, $\omega$ is
the circular frequency of the electromagnetic wave. Quantity ${\bf
n}^\times$ is called tensor dual to the vector ${\bf n}$. It results
in the vector product, if multiplied by a vector ${\bf a}$ as ${\bf
n}^\times {\bf a} = {\bf n} {\bf a}^\times = {\bf n} \times {\bf
a}$.

System (\ref{ODE1}) is the result of the variable separation in
Maxwell's equations. Eq. (\ref{ODE1}) includes two algebraic scalar
equations, therefore, two field components, $H_r$ and $E_r$, can be
expressed by means of the rest four components. This can be
presented as the matrix link between the total fields
$\hh=\hh_\rmt+H_r \er$ and $\ee=\ee_\rmt+E_r \er$ and their
tangential components $\hh_\rmt$ and $\ee_\rmt$:
\begin{eqnarray}
\left( \begin{array}{c} \hh (r) \\ \ee (r) \end{array} \right) = V
(r) \left( \begin{array}{c} \hh_\rmt (r) \\ \ee_\rmt (r) \end{array}
\right), \qquad &V = \left( \begin{array}{cc} I & \frac{
\sqrt{l (l+1)}}{\mu_{r}(r) k_0 r} \er \otimes \et \\
- \frac{\sqrt{l (l+1)}}{\varepsilon_{r}(r) k_0 r} \er \otimes \et &
I
\end{array} \right). \label{restore}
\end{eqnarray}
Excluding the radial components of the fields from Eq. (\ref{ODE1}),
we get to the system of ordinary differential equations of the first
order for the tangential components joined into the four-dimensional
vector $\ww(r)$ as
\begin{equation}
\frac{{\rm d} \ww(r)}{{\rm d} r}=\rmi k_0 M(r)\ww(r), \label{Meqt}
\end{equation}
where
\[ M=\left(\begin{array}{cc}A&B\\C&D\end{array}\right), \qquad
\ww=\left(\begin{array}{c}\hh_\rmt\\ \ee_\rmt\end{array}\right)
\equiv \left(\begin{array}{c} H_\theta \\ H_\varphi \\ E_\theta \\
E_\varphi \end{array}\right) ,
\]
\begin{eqnarray}
A = D = \frac{\rmi}{k_0 r} I,  \qquad
B=\varepsilon_t \er^\times - \frac{l (l+1)}{\mu_r k_0^2 r^2} \ep
\otimes \et, \qquad
C= - \mu_t \er^\times + \frac{l (l+1)}{\varepsilon_{r} k_0^2 r^2}
\ep \otimes \et. \label{ABCD}
\end{eqnarray}

Tangential field components $\hh_\rmt$ and $\ee_\rmt$ play the
important part, because they are continuous at the spherical
interface. Hence, they can be used for solving the scattering
problem. At first, we will analyze the situation of $r$-dependent
permittivities and permeabilities, which arise for the cloak
coatings of the spheres.

Excluding the $\varphi$-components of the fields from Eq.
({\ref{Meqt}), we derive the differential equation of the second
order for $w_\theta = \et \ww = (H_\theta, E_\theta)$:
\begin{equation}
w''_\theta + \frac{2}{r} w'_\theta - \left(\begin{array}{cc}
\frac{\varepsilon'_t}{\varepsilon_t} & 0 \\ 0 & \frac{\mu'_t}{\mu_t}
\end{array}\right) w'_\theta + \left[ k_0^2 \varepsilon_t \mu_t -
\frac{1}{r} \left(\begin{array}{cc}
\frac{\varepsilon'_t}{\varepsilon_t} & 0 \\ 0 &
\frac{\mu'_t}{\mu_t}\end{array}\right) - \frac{l (l+1)}{r^2}
\left(\begin{array}{cc} \frac{\varepsilon_t}{\varepsilon_r} & 0 \\ 0
& \frac{\mu_t}{\mu_r}
\end{array}\right) \right] w_\theta = 0, \label{wThetaInhom}
\end{equation}
where the prime denotes the $r$-derivative. Further we will apply
the condition on the medium parameters, which is usually used for
the cloaks: $\varepsilon_r(r) = \mu_r(r)$ and $\varepsilon_t(r) =
\mu_t(r)$. Then the equations for $H_\theta$ and $E_\theta$ coincide
and can be written in the form
\begin{equation}
w''_\theta + \left( \frac{2}{r} -
\frac{\varepsilon'_t}{\varepsilon_t} \right) w'_\theta + \left[
k_0^2 \varepsilon_t^2 - \frac{\varepsilon'_t}{r \varepsilon_t} -
\frac{l (l+1)}{r^2} \frac{\varepsilon_t}{\varepsilon_r} \right]
w_\theta = 0. \label{wThetaInhomCl}
\end{equation}

This equation can be solved analytically in the very few cases. As
an example, we can offer the inversely proportional transversal
$\varepsilon_t = a_1/r$ and radial $\varepsilon_r =a_2/r$ dielectric
permittivities. However, such dependencies do not provide the cloak
properties of the layer. Another analytical solution of Eq.
(\ref{wThetaInhomCl}) can be obtained for Pendry's cloak, that is
for the permittivities $\varepsilon_t =b/(b-a)$ and $\varepsilon_r =
\varepsilon_t (r - a)^2/r^2$.

In spite of analytical solutions cannot be found in the all required
situations, the general structure of solutions can be studied. The
solution of two differential equation of the second order
(\ref{wThetaInhom}) contains four integration constants $c_1$,
$c_2$, $c'_1$, and $c'_2$. The constants can be joined together to a
couple of vectors ${\bf c}_1 = c_1 \et + c'_1 \ep$ and ${\bf c}_2 =
c_2 \et + c'_2 \ep$. $\varphi$-components of the field vectors
$H_\varphi$ and $E_\varphi$ are expressed in terms of the already
determined $\theta$-components. The link between $\theta$- and
$\varphi$-components follows from Eq. (\ref{Meqt}). Summing up both
components, the resulting field can be presented as
\begin{equation}
\ww=S(r) {\bf C}, \quad S(r)=\left( \begin{array}{cc} \eta_1(r) &
\eta_2(r) \\ \zeta_1(r) & \zeta_2(r) \end{array} \right), \quad {\bf
C}= \left( \begin{array}{c} {\bf c}_1 \\ {\bf c}_2 \end{array}
\right), \label{WSC}
\end{equation}
where $\eta_1$, $\eta_2$, $\zeta_1$, and $\zeta_2$ are the
two-dimensional blocks of the matrix $S(r)$. Quantities $\eta_1$,
$\zeta_1$, and ${\bf c}_1$ correspond to the first independent
solution of Eq. (\ref{wThetaInhom}), while $\eta_2$, $\zeta_2$, and
${\bf c}_2$ do to the second independent solution. Therefore, the
general solution can be decomposed into the sum as $\ww = \ww^{(1)}
+ \ww^{(2)}$, where
\begin{eqnarray}
\ww^{(1)}= \left( \begin{array}{cc} \hh_{\rmt 1} \\
\ee_{\rmt 1} \end{array} \right) = \left( \begin{array}{cc} \eta_1  \\
\zeta_1 \end{array} \right) {\bf c}_1, \qquad
\ww^{(2)}= \left( \begin{array}{cc} \hh_{\rmt 2} \\
\ee_{\rmt 2} \end{array} \right) = \left(
\begin{array}{cc} \eta_2  \\ \zeta_2 \end{array} \right) {\bf c}_2.
\label{separatewaves}
\end{eqnarray}

Electric and magnetic fields of each independent wave are connected
by means of impedance tensor $\Gamma$ as $\ee_{\rmt j}=\Gamma_j
\hh_{\rmt j}$ ($j=1,2$). Thus, the impedance tensor equals
\begin{equation}
\Gamma_j(r)=\zeta_j(r) \eta^{-1}_j(r). \label{gen_Impedance}
\end{equation}

Vectors ${\bf c}_1$ and ${\bf c}_2$ can be expressed by means of the
known tangential electromagnetic field $\ww(a)$ as ${\bf C} =
S^{-1}(a) \ww (a)$. Then Eq. (\ref{WSC}) can be rewritten as follows
\begin{equation}
\ww(r)=\Om \ww(a), \qquad \Om=S(r) S^{-1}(a), \label{gen_evol_oper}
\end{equation}
where evolution operator (transfer matrix) $\Om$ connects tangential
field components at two distinct spatial points, $r$ and $a$.

Solution for the fields $\ee(\rr)$ and $\hh(\rr)$ can be written as
the sum over $l$ and $m$ of subsequent products of the tensor
$F_{lm}(\theta, \varphi)$ describing angle dependence (Eq.
(\ref{Flm})), matrix $V^l(r)$ restoring the fields with their
tangential components (Eq. (\ref{restore})), and tangential field
vectors (\ref{WSC}):
\begin{eqnarray}
\left( \begin{array}{cc} \hh (\rr)
\\ \ee(\rr) \end{array} \right)= \sum_{l=0}^\infty \sum_{m=-l}^l
\left( \begin{array}{cc} F_{lm} (\theta, \varphi) & 0
\\ 0 & F_{lm} (\theta, \varphi) \end{array} \right)
V^l(r) \left( \begin{array}{cc} \eta^l_1(r) & \eta^l_2(r)
\\ \zeta^l_1(r) & \zeta^l_2(r) \end{array} \right) \left(
\begin{array}{c} {\bf c}^{lm}_1 \\ {\bf c}^{lm}_2 \end{array}
\right). \label{gen_sol}
\end{eqnarray}

In general, the cloak solutions cannot be studied in the closed
form. Therefore, we will apply the approximate method of numerical
computations, which can be formulated as follows. Inhomogeneous
spherical shell $a<r<b$ is divided into $N$ homogeneous spherical
layers, i.e. replaced by the multi-layer structure. The number of
the layers strongly determines the accuracy of calculations. $j$th
homogeneous shell is extended from $a_{j-1}$ to $a_j$, where
$j=1,\ldots,N$, $a_{0}=a$ and $a_{N}=b$. Wave solution of the single
homogeneous layer can be presented in the form of evolution operator
$\Omega_{a_{j-1}}^{a_j}$. The solution for the whole inhomogeneous
shell is the subsequent product of the elementary evolution
operators, that is
\begin{equation}
\Omega_{a}^b =  \Omega_{a_{N-1}}^b \ldots \Omega_{a_{1}}^{a_2}
\Omega_{a}^{a_1}. \label{evolut_layers}
\end{equation}

The solution of equation (\ref{wThetaInhom}) with constant
permittivities $\varepsilon_r$, $\varepsilon_t$ and permeabilities
$\mu_r$, $\mu_t$ is expressed by means of the spherical functions.
In the layer, the general solution is represented using a couple of
independent solutions $g^{(1)}_{\nu}$ and $g^{(2)}_{\nu}$:
\begin{eqnarray}
\left( \begin{array}{c} H_\theta (r)\\ E_\theta (r) \end{array}
\right)
=\left( \begin{array}{c} g^{(1)}_{\nu_1}(k_t r) c_1 +
g^{(2)}_{\nu_1}(k_t r) c_2
\\ g^{(1)}_{\nu_2}(k_t r) c'_1 + g^{(2)}_{\nu_2}(k_t r)
c'_2 \end{array} \right), \label{theta-solution}
\end{eqnarray}
where $k_t = k_0 \sqrt{\varepsilon_t \mu_t}$, $\nu_1 = \sqrt{l (l+1)
\varepsilon_t/\varepsilon_r + 1/4} - 1/2$, $\nu_2 = \sqrt{l (l+1)
\mu_t/\mu_r + 1/4} - 1/2$. Functions $g^{(1,2)}_{\nu}$ of the order
$\nu$ can be spherical Bessel functions, modified spherical Bessel
functions, or spherical Hankel functions depending on the problem.

Blocks $\eta$ and $\zeta$ introduced in (\ref{WSC}) are the tensors
\begin{eqnarray}
&& \eta_{1,2}= g^{(1,2)}_{\nu_1} \et \otimes \et
- \frac{\rmi} {\mu_t k_0 r} \frac{{\rm d}(r
g^{(1,2)}_{\nu_2})}{{\rm d} r}  \ep \otimes \ep, \nonumber \\
&& \zeta_{1,2}= g^{(1,2)}_{\nu_2} \et \otimes \ep
+ \frac{\rmi} {\varepsilon_t k_0 r} \frac{{\rm d}(r
g^{(1,2)}_{\nu_1})}{{\rm d} r} \ep \otimes \et, \label{etazeta_ex}
\end{eqnarray}
which can be presented as two-dimensional matrices for computation
purposes.

Now we turn to the scattering of electromagnetic waves from the
two-layer sphere. The inner sphere of radius $a$ is homogeneous
isotropic one ($\varepsilon^{(1)}$ and $\mu^{(1)}$), while the
coating $a<r<b$ is characterized by the dielectric permittivity and
magnetic permeability tensors defined by Eq. (\ref{eps_mu}). We
suppose that an arbitrary electromagnetic field $\hh_{\rm inc}(\rr)$
and $\ee_{\rm inc}(\rr)$ is incident onto the two-layer spherical
particle from air ($\varepsilon^{(0)}=1$, $\mu^{(0)}=1$).

Wave solutions in each layer can be written using already known
general one (\ref{gen_sol}). Scattered field propagates in air and
can be presented in the form of superposition of diverging spherical
waves. Mathematically, such waves are described by spherical Hankel
functions of the first kind $h^{(1)}_\nu(x)$. Denoting the tensors
$\eta$ and $\zeta$ with Hankel functions $g^{(1)}_\nu=h^{(1)}_\nu$
as $\tilde{\eta}$ and $\tilde{\zeta}$, we get the scattered
electromagnetic field
\begin{eqnarray}
\left( \begin{array}{cc} \hh_{\rm sc} (\rr)
\\ \ee_{\rm sc}(\rr) \end{array} \right)= \sum_{l=0}^\infty \sum_{m=-l}^l \left(
\begin{array}{cc} F_{lm} & 0 \\ 0& F_{lm}
\end{array} \right)
V^l_{\rm sc}(r) \left(
\begin{array}{c} I \\ \tilde{\Gamma}^l(r) \end{array} \right) \tilde{\eta}^l(r)
(\tilde{\eta}^l(b))^{-1} \hh^{lm}_{\rm sc}(b), \label{scat_field}
\end{eqnarray}
where $\tilde{\Gamma}^l=\tilde{\zeta}^l (\tilde{\eta}^l)^{-1}$ is
the impedance tensor of the $l$th scattered wave, $\hh^{lm}_{\rm
sc}(b)$ is the tangential magnetic field at the particle interface
$r=b$.

Field inside the isotropic inner sphere is determined by the only
spherical Bessel function of the first kind. The field in the
inhomogeneous shell have no peculiarities and can be described in
terms of both spherical Bessel functions. Applying the evolution
operator $\Omega_{a}^r$ the electromagnetic field in the shell takes
the form
\begin{eqnarray}
\left( \begin{array}{cc} \hh_{{\rm sh}} (\rr)
\\ \ee_{{\rm sh}}(\rr) \end{array} \right) = \sum_{l=0}^\infty \sum_{m=-l}^l
\left( \begin{array}{cc} F_{lm} & 0 \\ 0& F_{lm}
\end{array} \right)
V^l_{{\rm sh}}(r) \Omega_{a}^r \left(
\begin{array}{c} I \\ \Gamma^l_1(a)
\end{array} \right) \hh^{lm}_{1}(a),
\label{inside_field}
\end{eqnarray}
where $\Gamma^l_1=\zeta^l_1 (\eta^l_1)^{-1}$ is the impedance tensor
of the $l$th wave inside the inner sphere, $\hh^{lm}_{1}(a)$ is the
tangential magnetic field at the inner interface of the shell $r=a$.

By projecting the fields onto the outer interface $r=b$ and
integrating over the angles $\theta$ and $\varphi$ with use of
orthogonality condition (\ref{orthogF}), we derive the boundary
conditions
\begin{eqnarray}
\ww_{\rm inc}^{lm}  +  \left( \begin{array}{c} I \\
\tilde{\Gamma}^l(b) \end{array} \right) \! \hh^{lm}_{\rm sc}(b)
= {\bf \Omega}_{a}^{b} \!\! \left( \begin{array}{c} I \\
\Gamma^l_1(a) \end{array} \right) \! \hh^{lm}_{1}(a),
\label{bc_final}
\end{eqnarray}
where
\begin{equation}
\ww_{\rm inc}^{lm} = \int_0^\pi \int_0^{2 \pi} \left(
\begin{array}{cc} F^+_{lm} (\theta, \varphi) I \hh_{\rm
inc} (b,\theta, \varphi)
\\ F^+_{lm} (\theta,\varphi) I \ee_{\rm
inc}(b,\theta, \varphi) \end{array} \right) \sin \theta {\rm d}
\theta {\rm d} \varphi. \label{Wlm}
\end{equation}

Expression (\ref{bc_final}) represents the system of four linear
equations for four components of the vectors $\hh_{\rm sc}^{lm}$ and
$\hh_{1}^{lm}$. Excluding the constant vector $\hh_{1}^{lm}$ we
derive the amplitude of the scattered electromagnetic field
\begin{eqnarray}
\hh^{lm}_{\rm sc} (b) = - \left[ \left( \begin{array}{cc}
\Gamma^l_1(a) & - I \end{array} \right) \Omega_{b}^{a} \left(
\begin{array}{c} I \\ \tilde{\Gamma}^l(b)
\end{array} \right) \right]^{-1}
\left[ \left( \begin{array}{cc} \Gamma^l_1(a) & - I
\end{array} \right) \Omega_{b}^{a} \ww_{\rm inc}^{lm}
\right], \label{c_sc}
\end{eqnarray}
where $\Omega_{b}^{a} = (\Omega_{a}^{b})^{-1}$.

Scattered field can be characterized by the differential
cross-section (power radiated in $\er$-direction per solid angle
${\rm d}o$)
\begin{equation}
\frac{{\rm d} \sigma}{{\rm d} o}=r^2 \frac{|\hh_{\rm
sc}(\rr)|^2}{|\hh_{\rm inc}(\rr)|^2}.
\end{equation}

In our notations, the differential cross-section averaged over the
azimuthal angle $\varphi$ (over polarizations) takes the form
\begin{equation}
\frac{{\rm d} \sigma}{\sin\theta {\rm d} \theta}=
\frac{b^2}{|\hh_{\rm inc}|^2} \sum_{m=-\infty}^\infty \left|
\sum_{l=|m|}^\infty F_{lm} (\theta, 0) \hh^{lm}_{\rm sc}(b)
\right|^2. \label{cross}
\end{equation}

From the point of view of the scattering theory, it is natural to
define the cloak as the specially matched layer that provides zero
scattering for any material inside. Such definition is based on the
main property of the cloak --- its invisibility (i.e. the cloak
cannot be detected by optical means). In \cite{} zero scattering was
proved analytically for the Pendry cloak.

Zero scattering is specified by the condition $\hh^{lm}_{\rm
sc}(b)=0$, which can be rewritten using Eq. (\ref{c_sc}) as follows
\begin{eqnarray}
\left( \begin{array}{cc} \Gamma^l_1(a) & - I
\end{array} \right) \Omega_{b}^{a} \ww_{\rm inc}^{lm}
=0. \label{hsc=0}
\end{eqnarray}
Arbitrary incident electromagnetic field $\ww_{\rm inc}^{lm}$ can be
excluded from this expression. In fact, zero scattering amplitude
can be obtained for the trivial situation: electromagnetic field is
scattered by the air sphere of radius $b$. This assertion can be
presented in the form analogous to Eq. (\ref{hsc=0}):
\begin{eqnarray}
\left( \begin{array}{cc} \Gamma^l_0(b) & - I
\end{array} \right) \ww_{\rm inc}^{lm}
=0, \label{hsc=0Air}
\end{eqnarray}
where $\Gamma^l_0$ is the impedance tensor of the $l$th wave in the
air sphere of radius $b$. Hence, we get to the relation
\begin{eqnarray}
\left( \begin{array}{cc} \Gamma^l_1(a) & - I
\end{array} \right) \Omega_{b}^{a}
=\left( \begin{array}{cc} \Gamma^l_0(b) & - I
\end{array} \right). \label{hsc=0CloakDef}
\end{eqnarray}

Impedance tensor $\Gamma^l_1$ contains material parameters
$\varepsilon^{(1)}$ and $\mu^{(1)}$ of the sphere inside the
cloaking shell. At the same time, zero scattering should be provided
for any $\varepsilon^{(1)}$ and $\mu^{(1)}$. It can be realized, if
the partial derivative of Eq. (\ref{hsc=0CloakDef}) on
$\varepsilon^{(1)}$ equals zero, that is $\left( \begin{array}{cc} I
& 0 \end{array} \right) \Omega_{b}^{a}=0$. By multiplying this
equation by $\Gamma^l_1$ and subtracting it from Eq.
(\ref{hsc=0CloakDef}), we get one more equation, which does not
contain the material parameters of the inner sphere: $\left(
\begin{array}{cc} 0 & I \end{array} \right) \Omega_{b}^{a}=
\left( \begin{array}{cc} -\Gamma^l_0(b) & I \end{array} \right)$.
Finally, we derive the evolution operator of the cloaking layer:
\begin{eqnarray}
\Omega_{b}^{a} = \left( \begin{array}{cc} 0 & 0 \\ \Gamma^l_0(b) & -
I \end{array} \right). \label{EvolOperCloak}
\end{eqnarray}

This condition defines the cloak and can be satisfied for the
specially chosen evolution operator $\Omega_{b}^{a}$ of the cloaking
shell. The evolution operator obtained is the degenerate block
matrix, which inverse matrix is not defined. It should be noted that
relation (\ref{EvolOperCloak}) is independent on the material of the
inner sphere. Per se, the derived relation connects the wave
solutions in the cloak (evolution operator $\Omega_{b}^{a}$) and
wave solutions in the homogeneous air sphere (impedance tensor
$\Gamma^l_0$), that is it performs the coordinate transformation for
the solutions, but not for the material parameters as usually.
Unfortunately, it is difficult to determine the dielectric
permittivity and magnetic permeability of the cloak from Eq.
(\ref{EvolOperCloak}).

Cloak condition $\hh^{lm}_{\rm sc}(b)=0$ substituted to Eq.
(\ref{bc_final}) results in expression
\begin{eqnarray}
\Omega_{b}^{a} \ww_{\rm inc}^{lm} =  \left( \begin{array}{c} I \\
\Gamma^l_1(a) \end{array} \right) \! \hh^{lm}_{1}(a).
\label{bc_final1}
\end{eqnarray}
Using (\ref{EvolOperCloak}) it is clear that $\hh^{lm}_{1}(a) = 0$
and $\ee^{lm}_{1}(a) \equiv \Gamma^l_1(a)\hh^{lm}_{1}(a) = 0$. So,
we may conclude that both electric and magnetic fields equal zero at
the boundary $r=a$, therefore, the electromagnetic field is equal to
zero at any spatial point inside the inner sphere. In the cloaking
shell, the fields equal zero at the inner boundary $r=a$ (owing to
the continuity of the tangent fields) and equal incident fields at
the outer boundary $r=b$. Then, the field inside the cloak is of the
form (see Eq. (\ref{inside_field}))
\begin{eqnarray}
\left( \begin{array}{cc} \hh_{{\rm sh}} (\rr)
\\ \ee_{{\rm sh}}(\rr) \end{array} \right) = \sum_{l=0}^\infty \sum_{m=-l}^l
\left( \begin{array}{cc} F_{lm} & 0 \\ 0& F_{lm}
\end{array} \right)
V^l_{{\rm sh}}(r) \Omega_{b}^r \ww_{{\rm inc}}^{lm}.
\label{inside_field1}
\end{eqnarray}

\section{Bell-Shaped Generating Function for Cloak Optimization}

Starting from this section we consider non-ideal cloaks since we use
a discrete model to compute and compare the far-field scattering. If
the ideal cloaks are considered, each of the cloak design is
equivalent leading to zero scattering. Realistic cloaks can be made
of multiple homogeneous spherical layers, which replace the
inhomogeneous cloaking shell. In this case the scattering is not
zero, but noticeably reduced, and such a cloak realization is called
non-ideal (see Fig. \ref{fig:0}). In this section we will find the
best non-ideal cloak providing the lowest cross-section among all
designs investigated.


\begin{figure}[t!]
\centering \includegraphics[width=4.7cm]{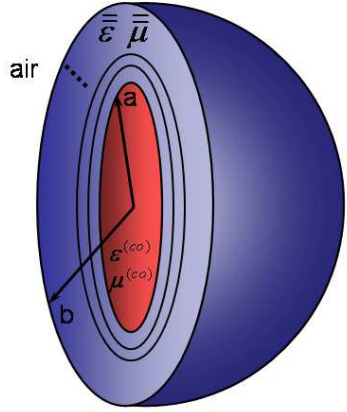}
\caption{[Color online] Illustration of the cloaking shell covering
the object to be concealed. We consider the spherical cloak in free
space with the inner radius $k_0 a=\pi$ and outer radius $k_0 b=2
\pi$. The core material is glass ($\varepsilon^{(co)}=1.45^2$ and
$\mu^{(co)}=1$). These quantities are used throughout the whole
paper. The material parameters $\overline{\overline{\varepsilon}}$
and $\overline{\overline{\mu}}$ are determined by applying the
proposed transformation-free method to an arbitrary cloaking
generating function. The cloaking shell is equally divided into $N$
layers (each layer is homogeneous and anisotropic), and the
scattering theory in \cite{MaHua} is used to compute the far-field
diagrams.} \label{fig:0}
\end{figure}


We will consider some typical generating functions (for transverse
dielectric permittivities) which exhibit different types of
profiles. The simplest generating functions are constant, linear,
and quadratic ones. Which of them provides the best cloaking
performance?

Constant generating function produces the dielectric permittivities
of the Pendry's classic spherical cloak as it has been demonstrated
in the previous section. Linear generating function can be generally
written as $g(r) = r - p$, where $p$ is a constant parameter. In
this case transverse and radial dielectric permittivities become
\begin{eqnarray}
\varepsilon_t (r) = \frac{ 2 b (r-p) }{(b-a) (b+a - 2 p)}, \\
\label{TDP_lin}
\varepsilon_r = \frac{ b (r-a)^2 (r+a - 2 p)^2 }{2 r^2 (r-p) (b-a)
(b+a - 2 p)}. \label{RDP_lin}
\end{eqnarray}
Parameter $p$ can take any value except $(a+b)/2$. It controls the
slope of the transverse permittivity function. If $p<(a+b)/2$,
$\varepsilon_t(r)$ linearly increases, and otherwise it
monotonically decreases.

Quadratic generating function has the general form
$g(r)=(r-p)(r-d)+s$, where $p$, $d$, and $s$ are tunable parameters.
The expressions for the permittivities in quadratic case can be
deduced to
\begin{eqnarray}
\varepsilon_t (r) = \frac{ b [(r-p)(r-d)+s] }{ P(b) }, \\
\label{TDP_quad}
\varepsilon_r = \frac{ b P^2(r) }{ r^2 [(r-p)(r-d)+s] P(b)},
\label{RDP_quad}
\end{eqnarray}
where
\begin{equation}
P(r) = \frac{r^3-a^3}{3} -(p+d)\frac{r^2-a^2}{2} + (pd + s) (r-a).
\end{equation}
Quadratic transverse permittivity is a parabola in graphical
presentation. The parabola can have a minimum (i.e., $s>s_0$) or
maximum (i.e., $s<s_0$), where $s_0 = -(b^2 + a b + a^2)/3 + (p+d)
(a+b)/2 - pd$.

Using these generating functions, some typical situations depicted
in Fig. \ref{fig:1} are presented. Profile 5 demonstrates the
permittivities for the constant generating function $g(r) = 1$
corresponding to Pendry's cloak. Linear generating functions are
presented in Profile 4 ($g(r)=r-a$) and Profile 6 ($g(r)=r-b$). The
other profiles are produced using quadratic generating functions.


\begin{figure}[t!]
\centering \includegraphics[width=10cm]{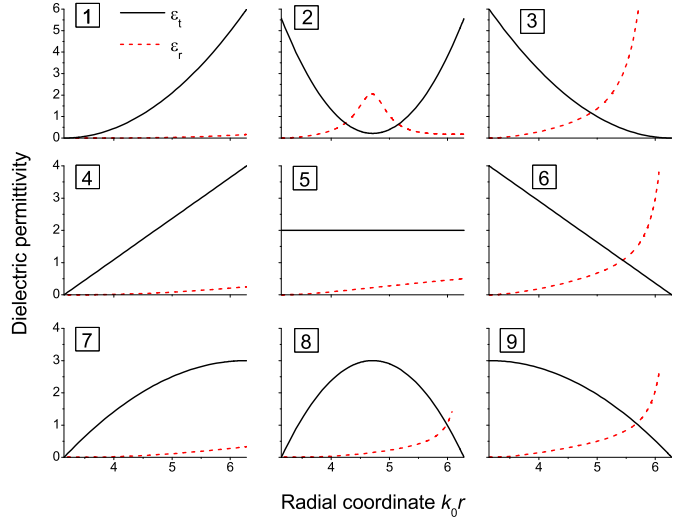} \vspace{-3ex}
\caption{[Color online] Transverse $\varepsilon_t$ and radial
$\varepsilon_r$ dielectric permittivities corresponding to different
profiles of generating function. Profiles 1--9 are described as
follows: (1) quadratic generating function with $p=0$, $d=b$,
$s=b^2/4$; (2) quadratic generating function with $p=a$, $d=b$,
$s=(b-a)^2/4+0.1$; (3) quadratic generating function with $p=a$,
$d=2b - a$, $s=(b-a)^2$; (4) linear generating function with $p=a$;
(5) constant generating function; (6) linear generating function
with $p=b$; (7) quadratic generating function with $p=a$, $d=2b-a$,
$s=0$; (8) quadratic generating function with $p=a$, $d=b$, $s=0$;
(9) quadratic generating function with $p=0$, $d=b$, $s=0$. }
\label{fig:1}
\end{figure}


The performances of different cloaks can be compared in terms of
their scattering cross-sections. The best cloaking design possesses
the lowest cross-section because of the reduced interaction of the
electromagnetic wave with the spherical particle. The inhomogeneous
anisotropic spherical cloaking shell is divided into $N$ homogeneous
anisotropic spherical layers. An experimental realization of this
multilayer cloak can be the sputtering onto the spherical core.
Throughout the whole paper we use $N=30$.

In Fig. \ref{fig:2} the total cross-sections resulting from
different generating functions are shown. Some profiles are
approximately equivalent, for example, 1, 2 and 3, or 4 and 6, or 7
and 9. Profiles 1--3 are characterized by concave-up transverse
dielectric permittivity $\varepsilon''_t>0$. According to Fig.
\ref{fig:2} they give rise to the worst results. The flat-curvature
profiles 4--6 characterized by $\varepsilon''_t=0$ are much better.
Pendry's cloak (number 5) stands out against the other
zero-curvature profiles. However, the most effective cloak design is
the case of concave-down transverse permittivity
$\varepsilon''_t<0$. Profiles 7, 8, and 9 are better than profiles
4, 5, and 6, respectively, by approximately 4.8 dB. The quadratic
cloak with concave-down transverse permittivity (bell-shaped cloak)
is shown to be the best candidate.


The maximum of $\varepsilon_t$ in profile 8 is in the middle
position of the cloaking shell. Shifting the maximum of such a bell
shape towards the limit at the outer boundary (i.e., profile 7) or
inner boundary (i.e., profile 9), the cloaking performance is
monotonically degraded as shown in Fig. \ref{fig:2}. If parameter
$s$ is extremely huge in quadratic generating function ($s
\rightarrow \infty$), the cloak permittivities coincide with those
of Pendry's cloak. Thus, the increase of $s$ improves the cloak 2
and deteriorates the cloak 8.


\begin{figure}[t!]
\centering \includegraphics[width=7cm]{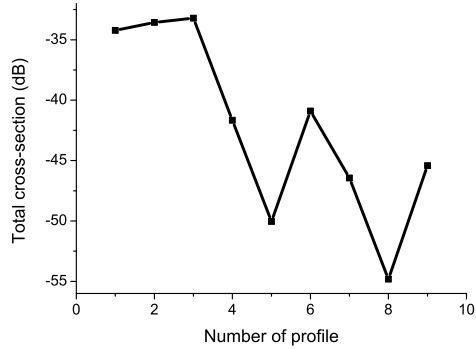} \vspace{-2ex}
\caption{ Total cross-sections for profiles of permittivities shown
in Fig. \ref{fig:1}. The number of discrete layers forming the cloak
equals $N=30$. } \label{fig:2}
\end{figure}


\section{The General Class of Bell-Shaped Cloaks}

From the previous section it is concluded that the bell-shaped
profile of the transverse dielectric permittivity leads to the
optimal non-ideal cloaking performance. In the present section we
will consider the general class of bell-shaped cloaks and choose the
best type.

Apart from the quadratic cloak, another three simple bell-shaped
profiles will be considered: Gaussian, Lorentzian, and Sech. All of
them have a single parameter $T$, which sets the width of the
profile. We take the maxima of such transverse permittivities in the
middle of the cloaking shell region (at the point $(a+b)/2$) to
compare with quadratic cloak.

Gaussian cloak has the generating function $g(r) = \exp [ -
(r-(a+b)/2)^2/(4 T^2)]$. The permittivity functions are
\begin{eqnarray}
\varepsilon_t &=& \frac{b}{2 \sqrt{\pi} T {\rm Erf} [(b-a)/(4 T)]}
{\rm e}^{-\frac{(r-(a+b)/2)^2}{4 T^2}} \nonumber \\
\varepsilon_r &=& \frac{ \sqrt{\pi} T b ({\rm Erf} [(b+a-2 r)/(4 T)]
- {\rm Erf} [(b-a)/(4 T)])^2}{2 r^2 {\rm Erf} [(b-a)/(4 T)]} {\rm
e}^{\frac{(r-(a+b)/2)^2}{4 T^2}}. \label{cloak_Gauss}
\end{eqnarray}

The generating function of the Lorentzian cloak is  $g(r) = 1/[1+
(r-(a+b)/2)^2/T^2]$. The transverse and radial permittivities for
this cloak are of the form
\begin{eqnarray}
\varepsilon_t &=& \frac{b}{2 T [1+ (r-(a+b)/2)^2/T^2] \arctan [(b-a)/(2T)]} \nonumber \\
\varepsilon_r &=& \frac{ T b (\arctan [(b+a-2 r)/(2 T)] - \arctan
[(b-a)/(2 T)])^2}{2 r^2 \arctan [(b-a)/(2 T)]}\times \nonumber \\
&&\left(1+ \frac{(r-(a+b)/2)^2}{T^2} \right). \label{cloak_Lorentz}
\end{eqnarray}

Sech cloak generating function depends on the radial coordinate as
$g(r) = {\rm sech}^2 [(r-(a+b)/2)/T]$. The permittivities are as
follows
\begin{eqnarray}
\varepsilon_t &=& \frac{b {\rm sech}^2 [(r-(a+b)/2)/T]}{2 T \tanh[(b-a)/(2T)]} \nonumber \\
\varepsilon_r &=& \frac{ T b (\tanh [(2 r - b - a)/(2 T)] - \tanh
[(b-a)/(2 T)])^2}{2 r^2 {\rm sech}^2 [(r-(a+b)/2)/T] \tanh [(b-a)/(2
T)]}. \label{cloak_Sech}
\end{eqnarray}

We choose equal 3 dB bandwidths for various transverse permittivity
profiles to compare different cloaks. Parameters $T$ which are tuned
to provide identical 3 dB bandwidth for each cloak are given in the
caption of Fig. \ref{fig:3}. In this figure we show the total
cross-sections of quadratic, Gaussian, Lorentzian, and Sech cloaks.
Profiles of Gaussian, Lorentzian, and Sech cloaks are very close,
resulting in similar scattering cross-sections. The influence of the
permittivity functions on the cloak performance is difficult to tell
among these three cloaks. However, it is shown that quadratic cloak
in Fig. \ref{fig:3} provides better invisibility when its transverse
permittivity vanishes at the inner and outer boundaries of the
cloaking shell.


\begin{figure}[t!]
\centering \includegraphics[width=8cm]{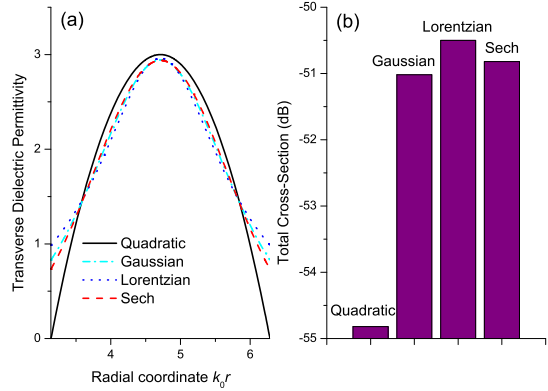}
\vspace{-2.3ex}\caption{[Color online] (a) Profiles of transverse
dielectric permittivity for quadratic (profile No. 8 in Fig.
\ref{fig:1}), Gaussian, Lorentzian, and Sech cloaks and (b) total
cross-sections of these cloaks. Parameter $T$ equals $(b-a)/4
\sqrt{2 \ln 2}$ for Gaussian, $(b-a)/(2\sqrt{2})$ for Lorentzian,
and $(b-a)/(2 \sqrt{2} \ln(\sqrt{2}+1))$ for Sech cloak. The number
of discrete layers forming the cloak equals $N=30$. } \label{fig:3}
\end{figure}


Since the shapes of the Gaussian, Lorentzian, and Sech cloaks are
similar, we can just select one of them, (e.g., Gaussian) to
investigate the significance of the profile, which can be varied by
the parameter $T$. The results are demonstrated in Fig. \ref{fig:4}.
The total cross-section has a minimum, which does not provide better
cloaking than quadratic one though. The cross-section minimization
is achieved approximately at $T=0.3 a$. This profile is shown in
Fig. \ref{fig:4}(a) along with profiles for other $T$ parameters.
According to this figure the minimization profile has the 3 dB
bandwidth equal $(b-a)/2$. Such a profile is neither too narrow nor
too wide because narrow profiles ($T\rightarrow 0$) need extremely
high discretization and wide profiles ($T\rightarrow \infty$) tend
to the limit of Pendry's cloak as shown in Fig. \ref{fig:4}(b).

Thus the bell-shaped quadratic cloak is preferred for non-ideal
cloak design, which has the lowest cross-section among all
bell-shaped cloaks considered in this section. In the following
section we will show how the quadratic cloak results can be
improved.


\begin{figure}[t!]
\centering \includegraphics[width=8cm]{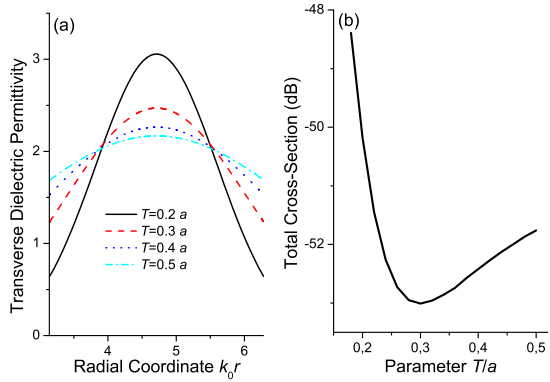}
\vspace{-2.5ex} \caption{[Color online] (a) Profiles of transverse
dielectric permittivity for Gaussian cloaks with different
parameters $T$ and (b) total cross-sections versus parameter $T$.
The number of discrete layers forming the cloak equals $N=30$.}
\label{fig:4}
\end{figure}


\section{Improved Quadratic Cloaks}

Quadratic cloak is characterized by very simple profile of the
transverse dielectric permittivity. Also, the quadratic cloak has
the scattering almost 5 dB lower than that of classic spherical one.
Our aim of this section is to find a way of creating the
high-performance cloaks based on the transformation-free design
method and bell-shaped quadratic cloak. The high-performance cloak
should be similar to the quadratic one. Transverse permittivity
should have a maximum and vanish at the inner and outer radii of the
shell: $\varepsilon_t(a)=\varepsilon_t(b)=0$. These properties can
be satisfied for general generating function of the form
\begin{equation}
g(r)=(r-a)(r-b)g_1(r). \label{genf1}
\end{equation}

By choosing function $g_1(r)$, we can set the permittivity profile
of the cloak. The function $g_1(r)$ can take arbitrary values at the
cloak edges $r=a$ and $r=b$, though it should provide the maximum of
the transverse permittivity. At first we will consider the maximum
at the center of the cloak $r=(a+b)/2$, and then the effect of the
non-central maximum position will be studied. For instance function
$g_1(r)$ can be selected with Gaussian profile. Then the cloak can
be called Gaussian-quadratic one. However, such a design is worse
than the simple quadratic shape. To provide the better design we
will focus the quadratic dependence using the $g_1(r)$ function
\begin{equation}
g_1(r)=((r-p)(r-d)+(d-p)^2/4+s)^n. \label{genf_pow}
\end{equation}
When $n=0$, it is just the bell-shaped quadratic cloak discussed
before. The permittivities at $n>0$ are suppressed due to lengthy
expressions.

\begin{figure}[t!]
\centering \includegraphics[width=8cm]{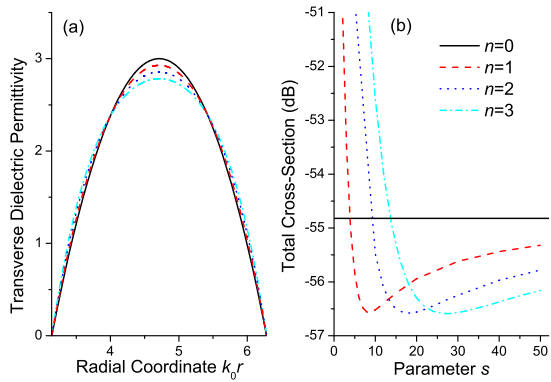}
\vspace{-2.5ex} \caption{[Color online] (a) Profiles of transverse
permittivity for power quadratic cloaks and (b) total cross-sections
of these cloaks versus parameter $s$. Parameters: $p=a$, $d=b$,
$N=30$. In (b), only $s>0$ is considered for cloaking purposes
because the total cross-sections corresponding to $s<0$ are
significantly larger.} \label{fig:5}
\end{figure}

In the generating function set by Eqs. (\ref{genf1}) and
(\ref{genf_pow}) we can vary the power term $n$ (the curvature of
the transverse permittivity profile at peak), parameters $s$ (the
deviation from the quadratic cloak) and $d$ (the deviation of the
permittivity peak from the center of the cloaking region). At $n=0$
the generating function is independent on $s$ and $d$ so the total
cross-section is the straight line in Fig. \ref{fig:5} (the solid
line), where other positive values of $n$ are shown as well. The
minima of the cross-sections (the best cloaking performance) occurs
to the parameter $s$ approximately at $s_{min}\approx 9 n$, i.e.,
linear to the power $n$. At larger parameter $s$, the curves tend to
the cross-section of the quadratic cloak. At small and negative $s$,
the shape of the transverse permittivity contains the minimum and a
couple of maxima, therefore the total cross-section is substantially
increased. In Fig. \ref{fig:5}(b), the cloaking performance is
obviously improved compared with the quadratic cloak. Let us further
study the effect of the peak position of the profile, which is
controlled by the parameter $d$.


The case $d=b$ describes that the position of the permittivity
maximum is in the center of the cloaking shell region. If $d<b$
($d>b$), the maximum is shifted towards the outer (inner) radius of
the cloaking shell. Fig. \ref{fig:6} shows that the central position
of the permittivity maximum is not the optimal choice. The
minimization of the cross-section is achieved for $d \approx 0.84
b$. Such a non-central position is expected to result from the
spherically curvilinear geometry of the cloak.


\begin{figure}[t!]
\centering \includegraphics[width=8cm]{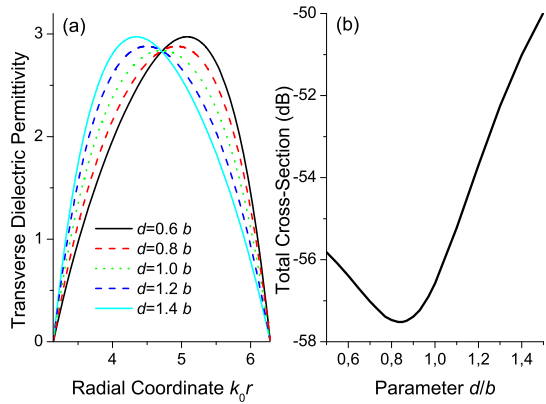}
\vspace{-2.5ex} \caption{[Color online] (a) Profiles of transverse
dielectric permittivity for power quadratic cloaks and (b) total
cross-sections of these cloaks vs. parameter $d$. Parameters: $p=a$,
$n=2$, $s=18$, $N=30$. } \label{fig:6}
\end{figure}


Compared with the quadratic cloak, the improvement of the
performance of power quadratic cloak is considerable: the total
cross-section is further decreased from $-54.84$ dB to $-57.52$ dB.
The improvement is caused by the shape of the profile. The profile
should be parabolic-like with a slightly deformed shape.

It is also important to consider the differential cross-sections
which provides the scattering intensity at an arbitrary angle. In
Fig. \ref{fig:7} we show the differential cross-sections for some
typical cloaking designs designed by the transformation-free method
and considered in a non-ideal situation. The common feature of the
cloaks is the reduced backscattering. It is seen that the classic
spherical cloak is the most visible one. The quadratic cloak (blue
dotted line) can provide much lower scattering over almost all
angles compared with Pendry's and Gaussian's bell-shaped cloak. The
power quadratic cloak is able to further bring down the scattering
of the quadratic cloak near the forward direction.


\begin{figure}[t!]
\centering \includegraphics[width=8cm]{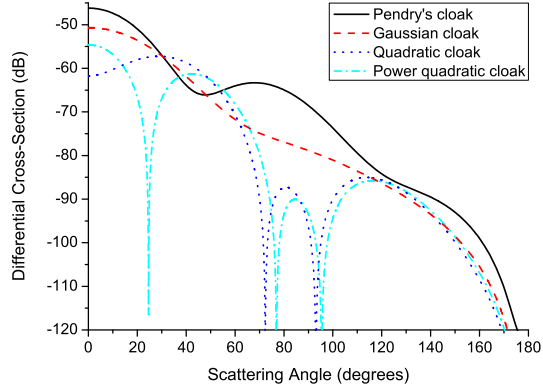}
\vspace{-2.5ex}\caption{[Color online] Differential cross-sections
of the cloaks derived by the proposed transformation-free method.
The cloak parameters of each given design have been selected to
provide the best performance respectively. Parameters: $T=0.3 a$ for
Gaussian cloak; $s=18$, $n=2$, $p=a$, and $d=0.84 b$ for power
quadratic cloak; $N=30$. } \label{fig:7}
\end{figure}

\begin{figure}[t!]
\centering
\includegraphics[width=8cm]{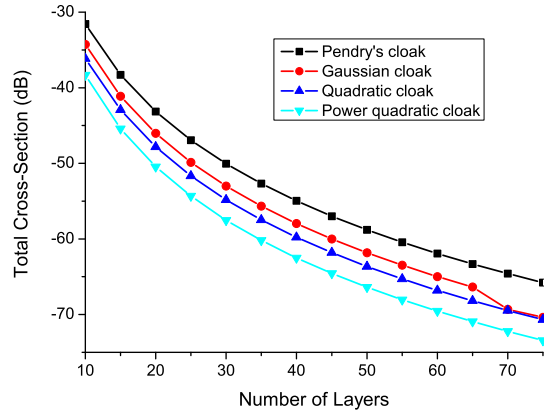}\vspace{-2.5ex} \caption{
[Color online] Total cross-sections of different cloak designs
versus the number of spherical layers $N$. Parameters of the cloaks
are the same as those in Fig. \ref{fig:7}. } \label{fig:8}
\end{figure}
However, one may question that our non-ideal situation may approach
to the ideal case when the discretization is high (i.e., $N$ is much
larger than 30). If so, each cloak derived from our proposed
reversed algorithm should be more and more identical to each other.
Theoretically, it is true provided that $N\rightarrow \infty$, while
the influence of the disretization number $N$ on the optimization
result is still of significant importance in practice. We calculate
the total cross-sections for different numbers of spherical layers
in Fig. \ref{fig:8}. In general, we observe the conservation of our
conclusions on the optimization for $N=30$, except for that the
performance of the Gaussian cloak matches with that of the quadratic
cloak at $N=70$. When $N$ is small, the dependence of scattering
reduction on the value of $N$ is nonlinear. For great number of
layers $N$ the curves become mostly linear as shown in Fig.
\ref{fig:8}.


\section{Conclusion}

It has been found that the bell-shaped cloaks provide the smallest
interaction of the cloaking shell with the electromagnetic radiation
under the non-ideal situation (i.e., the cloaking shell is
discretized into $N$ layers). Among the bell-shaped cloaks, we have
compared quadratic, Gaussian, Lorentzian, and Sech cloaks. The last
three are very similar in profile shape and dependence of
controlling parameters. We have concluded that the best performance
is achieved when the bell-shaped transverse permittivity profiles
which vanish at the inner and outer radii of the cloaking shell. The
simplest design of such a type is the quadratic cloak. Improved
invisibility performance can be provided by the power quadratic
cloak with the maximum of the permittivity profile slightly shifted
towards the outer boundary. The decrease of cloak's overall
scattering is about 7.5 dB compared with the classical Pendry's
design, and the improvement is steady even when the discretization
$N$ is quite high.

\section*{Acknowledgement}
This research was supported in part by the Army Research Office
through the Institute for Soldier Nanotechnologies under Contract
No. W911NF-07-D-0004. A. Novitsky acknowledges the Basic Research
Foundations of Belarus (F08MS-06). We thank Prof. John Joannopoulos
and Prof. Steven Johnson for their stimulating comments and
revisions throughout the manuscript preparation.



\end{document}